\documentstyle{article}
\input FEYNMAN

\begin{document}

\begin{flushright}
GUTPA/01/08/01 \\
DCPT-01/63\\
\end{flushright}

\begin{center}

{\Large \bf 
Quantum Numbers of AGUT Higgs Fields from the Quark-Lepton Spectrum}

\vspace{20pt}

{\bf C.D. Froggatt}

\vspace{6pt}

{ \em Department of Physics and Astronomy,\\
University of Glasgow, Glasgow, G12 8QQ, UK\\}

\vspace{12pt}

{\bf H.B. Nielsen}

\vspace{6pt}

{\em Niels Bohr Institute, \\
Blegdamsvej 17-21,
DK 2100 Copenhagen, Denmark\\}

\vspace{6pt}

and\\

\vspace{6pt}
{\bf D.J. Smith}

\vspace{6pt}

{\em Department of Mathematical Sciences,\\
University of Durham, Durham, DH1 3LE,UK\\} 

\vspace{10pt}
\end{center}
\begin{abstract}
A series of Higgs field quantum numbers in the anti-grand 
unification model, based on the gauge group $SMG^3 \times U(1)_f$, 
is tested against the spectrum of quark and lepton masses and 
mixing angles. A more precise formulation of the statement that 
the couplings are assumed of order unity is given. It is found that 
the corrections coming from this more precise assumption do not 
contain factors of order of the number of colours, $N_c= 3$, as one 
could have feared. We also include a combinatorial correction factor, 
taking account of the distinct internal orderings within the chain 
Feynman diagrams in our statistical estimates.

Strictly speaking our model predicts that the uncertainty in its predictions 
and thus the accuracy of our fits should be $\pm 60\%$. Many of 
the best fitting quantum numbers give a higher accuracy fit 
to the masses and mixing angles, although within the expected fluctuations 
in a $\chi^2$.  This means that our fit is as good as it can possibly be.

\end{abstract}

\thispagestyle{empty} \newpage

\section{Introduction \label{intro}}

Almost the only window of information 
available to glimpse the laws of nature 
beyond the Standard Model is at present provided by the ca 20 parameters, 
mainly already measured but not predicted by the Standard Model itself.
Among these, the 13 parameters comprising the  masses, 
mixing angles and CP violating phase for the fermions---quarks and 
leptons---make up more than half.  
We have earlier put forward a model \cite{PreviousPapers} 
as a candidate for explaining 
the orders of magnitudes of the masses for the six quarks and the three 
charged leptons. In this anti-grand unification theory (AGUT) 
the smallness of most of the Standard Model
Yukawa couplings is to be understood as a consequence of there 
existing some approximately conserved gauge charges forbidding these 
Yukawa couplings \cite{FN}. 
We actually managed in our AGUT model
to successfully fit all the 9 charged particle masses\footnote{We do not 
discuss the neutrino mass problem here, which requires the 
introduction of another mass scale and corresponding Higgs 
particle.} and the three mixing angles 
using only three Higgs field vacuum expectation values, in addition 
to the usual Weinberg Salam Higgs field. 

In the present article we want to put forward a few refinements of this 
model: 

1) First we want to take into account that the number of different Feynman 
diagrams that can contribute to a given mass matrix element---a given
effective Higgs-Yukawa coupling---becomes rather large when the diagram 
is complicated, because it involves many 
Higgs field-lines.
This number of diagrams can be expressed in terms of 
factorials and can easily be so
large that, even for order of magnitude calculations, we should include them.

2) In our previous work on this AGUT-fermion mass model, we introduced 
a  Higgs field $S$ for which we could take the vacuum expectation 
value (VEV) in the fundamental (Planck) units to be of order 
unity. Inclusion of the factorial corrections, just mentioned under point 1), 
actually makes the 
approximation of the $S$ field VEV by unity give a poorer fit to 
the data. So, in order to regain as good a fit as before, 
we are forced to allow the VEV of 
the Higgs field $S$ to be a variable parameter when 
the factorials are included; although its fitted 
value still turns out to be rather close to unity.

3) If there are many types of particles in the model, or if we have 
particles that like quarks are say triplets under colour 
symmetry\footnote{In the AGUT model there is even a different colour 
group for each proto-generation.},
one would a priori imagine that the number of particles or the number
of colours---i.e. 3---were tacitly treated as being of order unity
in our previous fits of the quark and lepton masses. If this were 
really so, one would expect corrections of the order of, say, a factor 
3 in our predictions. Actually our previous fits 
\cite{PreviousPapers} always agreed 
with data to better than a factor 3, except for perhaps the most 
sensitive of our predictions---the CP-violation strength---which  
is especially uncertain, because it is a product of very many only order of
magnitudewise known factors; in fact it was not 
included in the fit but simply
predicted by our model. Even this deviation of our CP-violation prediction 
from the experimental number is only just about a factor 3. 
In order to make sense of such a good agreement for our 
previous fits, we have to formulate 
a more precise form of the loose statement that the coupling
constants are all of order unity at the Planck scale; so that 
we work to an accuracy where the number of colours (for the three  
SU(3) gauge groups in the model) and the number of species of particles, 
with masses of order the Planck scale, 
become large compared to the 
``order of magnitude accuracy'' to which we hope to fit. We must 
hope---and indeed the estimates in section 6
suggest the hope is fulfilled---that we do not get large
factors depending on the number of Planck scale 
species postulated in the model, because we do not know how many they 
should be. Rather we must show that their number is not important,
provided they are at least there at all.

It is the purpose of the present article to include the two
first improvements 1) and 2), and to discuss the third potential
correction 3) which, however, turns out to actually only give somewhat 
milder corrections. For example we show that our prediction
of order of magnitude degeneracy for quark and lepton masses in the same
generation (except for the $t$ and $c$ quarks) remains valid, without any 
order $N_c=3$ factors. So it becomes understandable that we can get 
order of magnitude fits with only about 40\% deviations from the data. 

Once we embark on including factors of suppression connected with our 
Higgs field $S$, which were previously set equal to unity, 
it matters how many times the nonzero VEV of this field $S$
is needed to generate a given mass matrix element (or effective Yukawa 
coupling). The required number of factors of this VEV in turn depends 
on the gauge quantum numbers of the other AGUT
Higgs fields, which we called $W$, $T$ and $\xi$. So their quantum 
numbers must now be specified completely and not just 
modulo the quantum numbers of our Higgs field $S$, as was sufficient 
as long as the VEV of the $S$ field 
could be considered to be of order unity. 
This specification of the precise gauge quantum numbers
for the Higgs fields $W$, $T$, $\xi$ and the field $\Phi_{WS}$,
containing the phenomenological 
Weinberg Salam Higgs field, 
gives rise to an infinite number of models. However, one should
first of all remark that it is extremely reasonable to assume that, in 
a specific version of the model, the charges can only take on 
discrete---essentially integer in appropriate units---values, 
and secondly we do not expect them to be very large.
So, at the end, the number of models to be considered is rather limited.
We shall label the various considered models by a set of
integers $(\alpha, \beta, \gamma, \delta)$ defined below 
in section \ref{higgs}. 

In the following section \ref{agut} 
we give a short review of the AGUT model.
Then, in the next section \ref{fac}, we 
explain and implement the factorial correction coming from  
point 1): the large numbers of permutations of the Higgs fields
in the Feynman diagrams. In section \ref{higgs} we then define the set of 
models obtained by the various specific quantum number choices for
our Higgs fields, i.e.~by choosing $(\alpha, \beta, \gamma, \delta)$.
In section \ref{results} we present the  
results of fitting quark and lepton masses and mixing angles 
with the various parameter choices and with the factorial 
corrections included. Also dicussed in this
section are two variants of how one interprets the order of unity factors
in the calculation of the predicted masses etc.: one can either just
put random phase factors on the products of the suppression factors and 
the factorial corrections, or one can further 
provide them with a random factor 
of order unity. The latter variation is not so 
important when one term dominates but when, as for the case of the 
d-quark mass in our model, more than one term contributes, it may 
give a correction. The best fitting 
Higgs field quantum number combinations are selected. 
Then follows, in section \ref{unity}, our proposal for making the 
concept of the couplings being of order unity more precise.
The outcome of this discussion is really that the results 
of section \ref{results} are not significantly modified 
by the more careful definition of
what the couplings being of order unity means.
In section \ref{deviation} we present  
the argument that, since many Feynman diagrams contribute, 
each mass matrix element in our model should have a 
Gaussian distribution in the complex plane. 
This, in turn, has the significance that 
we can even calculate an expectation
for what the random element in the masses should be and, thus, predict 
the degree of deviation between our fit and the experimental numbers.
Indeed we shall see that our prediction for the uncertainty agrees 
well with the experimental deviation of the fit.
Finally in section \ref{conclusion} we present our conclusions.      
 
\section{Our AGUT-model, a short review \label{agut}}

Anti-grand unification is a well suited name for our AGUT model 
in a number of senses, as we will now explain. It is similar to 
the usual grand unified theories (GUTs), which are based on $SU(5)$, 
$SO(10)$ or some other simple group, in the sense that we postulate 
the extension of the Standard Model Group \cite{ORaif}, $SMG$ = 
$ S(U(2)\times U(3))\approx U(1) \times SU(2)\times SU(3) $, 
to a larger gauge group at a very high energy of order 
the Planck scale. However 
the AGUT gauge group can be considered ``anti'', as it is not at all 
simple in the mathematical sense of the word. On the contrary, 
its Lie algebra consists altogether of ten cross product factors:
four $U(1)$ factors, three $SU(2)$ factors and three $SU(3)$ factors. 
Below the AGUT scale there is no supersymmetry nor other new 
physics beyond the Standard Model.

Further the usual GUTs unify by combining several Standard Model Group 
quark and lepton irreducible representations into the same 
irreducible representation of the grand unified group. Our AGUT is 
``anti'' in the sense that it is characterised by not uniting any of the 
Standard Model Group irreducible representations into a larger 
irreducible representation. In this way our model avoids the 
exact mass degeneracies predicted from the minimal $SU(5)$ GUT, 
according to which the running masses of the quarks with charge 
$-1/3$ are degenerate with their charged lepton partners at the 
unification scale. Honestly speaking, except popssibly for the case 
of the $b$ quark being degenerate with the $\tau$ lepton, 
these predictions are wrong. In fact the unwanted predictions,
$m_s = m_{\mu}$ and $m_d = m_e$, can only be tolerated 
by complicating the $SU(5)$ model \cite{GJ}. However it turns out that 
the AGUT model replaces these exact $SU(5)$ predictions by only order 
of magnitude degeneracy predictions, which are compatible with experiment. 

While $SU(5)$ is the smallest simple group 
unifying quarks and leptons, the AGUT group
$SMG^3\times U(1)_f$ is the biggest group, which keeps the 
Standard Model irreducible representations separate, in the 
following sense: there is no larger group, containing $SMG^3\times U(1)$ 
as a subgroup, which is faithfully represented on the Standard Model 
quark and lepton fields alone (no right-handed neutrinos or 
other new fermions), without gauge or mixed anomalies.
Indeed the $SMG^3\times U(1)$ group is uniquely specified as the 
group $G$ extending the Standard Model group and satisfying the 
following four postulates \cite{trento}.
\begin{enumerate}
\item $G$ should transform the three generations of Standard Model 
Weyl particles into each other unitarily, so that $G \subseteq U(45)$.
\item $G$ should be anomaly free, even without using the 
Green-Schwarz \cite{green} anomaly cancellation mechanism.
\item The fifteen irreducible representations of the Standard Model 
Weyl fields remain irreducible under $G$.
\item $G$ is the maximal group satisfying the other three postulates.
\end{enumerate}

Also the Standard Model gauge coupling constants do not unify 
in the AGUT model, but their values have been successfully 
predicted, at the Planck scale, using the so-called multiple 
point principle \cite{glasgow}. According to this principle, 
the coupling constants are fixed at such values as to 
ensure the existence of many vacuum states with the same energy 
density. We note that the top quark mass 
and the Weinberg Salam Higgs particle mass can also be 
predicted using this principle \cite{tophiggs}. 

Really the existence of this AGUT group means that,
near the Planck scale, each of the three quark-lepton 
proto-generations has its own $SMG$ gauge 
group and associated 12 gauge particles---i.e. the gauge 
bosons also come in generations. In addition there is an 
extra abelian $U(1)_f$ gauge boson; the corresponding gauge 
charge $Q_f$ is not carried by the left-handed fermions 
or any of the first proto-generation particles. So the 
non-zero $U(1)_f$ quantum numbers for the proto- quarks and 
leptons can be chosen as follows:
\begin{equation}
Q_f(\tau_R) = Q_f(b_R) = Q_f(c_R) = 1
\end{equation}
\begin{equation}
Q_f(\mu_R) = Q_f(s_R) = Q_f(t_R) = -1.
\end{equation}
We stress that the physical quarks and leptons are superpositions of 
the proto-fermions. In particular it turns out that the right-handed 
charm and top quarks are essentially permuted: the physical 
right-handed top quark is the Weyl component with second 
proto-generation quantum numbers, namely the right-handed proto-charm 
quark, to first approximation. 

The representations of the Standard Model Group 
$SMG = S(U(2) \times U(3))$ satisfy the charge quantization rule:
\begin{equation}
y/2 + d/2 + t/3 =0 \qquad ( \mbox{mod} \ 1)
\label{SMGiChQu}
\end{equation}
Here $y$ is the conventional weak hypercharge and $d$ is the duality,
which is defined to be $0$ when the weak isospin is integer and 
$d=1$ when it is half integer.  
Similarly the triality $t$ is defined to be $0$ for an $SU(3)$ colour 
singlet and $t= \pm 1$ for a triplet or antitriplet respectively. 
In the AGUT model this charge quantisation rule is satisfied 
separately for the $SMG$ quantum numbers associated with each 
quark-lepton generation.

Now we turn to the breaking of the AGUT gauge symmetry group 
$SMG^3 \times U(1)_f$ down to the Standard Model group SMG, 
which is supposed to occur close to the Planck scale. 
Unlike for the quarks and leptons, the gauge quantum numbers of the 
Higgs fields responsible for this symmetry breaking are not 
determined by the theoretical structure of the model. We 
assume that these Higgs fields have ``small'' quantum numbers, 
belonging to singlet or fundamental representations of the 
non-Abelian groups. The charge quantisation rule, Eq.~(\ref{SMGiChQu}), 
then determines the non-Abelian quantum numbers from the Abelian 
ones. So it is sufficient to specify just the Abelian quantum numbers 
in the form of a $U(1)$ charge vector:
\begin{equation}
{\bf Q} \equiv \left( \frac{y_1}{2}, \frac{y_2}{2}, \frac{y_3}{2}, 
Q_f \right) 
\end{equation}
where $y_i$ is the conventional weak hypercharge for the $i$'th 
proto-generation. For example the Higgs field $S$, introduced in 
\cite{PreviousPapers} with a VEV of order unity in fundamental units, 
has the Abelian quantum numbers ${\bf Q}_S = \left( \frac{1}{6}, 
-\frac{1}{6}, 0, -1 \right)$. It follows from the charge quantisation rule 
that the full set of AGUT quantum numbers for the $S$ field are: 
$\left({\bf 3},{\bf 2},\frac{1}{6};
{\bf \overline{3}}, {\bf 2},-\frac{1}{6};{\bf 1},{\bf 1},0;-1\right)$,
where the sets of three 
quantum numbers specifying the representations under the three 
proto-generation $SMG$-groups are separated by semi-colons.

As long as we consider the VEV of the Higgs field $S$ to be 
of order unity, we are in reality working with the 
group to which $S$ breaks the AGUT group down. Then it is only 
necessary to specify the quantum numbers under this subgroup 
(i.e.~modulo the quantum numbers of $S$). Thus the quantum numbers 
${\bf Q}_T = \left( 0, -\frac{1}{6}, \frac{1}{6}, -\frac{2}{3} \right)$ 
for the Higgs field $T$ used in \cite{PreviousPapers} could 
just as well be taken to be  
${\bf Q}_T = \left( -\frac{1}{6}, 0, \frac{1}{6}, \frac{1}{3} \right)$ 
modulo the $S$ quantum numbers. This transformed 
set is of course arbitrary as far as adding $S$ quantum numbers is 
concerned, but we have chosen it as in some sense the smallest 
Abelian quantum number choice. 

Another crucial assumption in our model is the existence,  
at the fundamental (Planck) scale, of a large spectrum 
of vector-like Dirac particles with the quantum numbers 
needed in the Feynman diagrams generating the quark-lepton 
mass matrices \cite{FN}. Furthermore it is assumed that all the couplings, 
allowed by the AGUT gauge symmetries, of these Planck scale particles 
to each other and to the lighter Standard Model particles are of order 
unity.

In practice we construct the mass matrix elements 
between the known Standard Model left- and right-handed Weyl fields 
one by one, by finding out first which 
quantum numbers need to be broken to make a given 
matrix element different from zero. Knowing the needed breaking 
quantum numbers, we then ask which combination of Higgs fields 
in our model can together cause that breaking. Measuring 
everything in ``fundamental'' units we can say that, 
since everything is of order unity 
except the small VEVs, we just take the product of the VEVs for the 
needed combination of Higgs fields. Finally the mass matrix element 
in question is simply order of magnitudewise equal to this product. 
However we should mention that there is an ambiguity  
concerning which combination
of Higgs fields to use, since the Abelian quantum numbers of our proposed 
Higgs fields $W$, $T$ and $\xi$ obey a linear relation:
$3 {\bf Q}_W  -9{\bf Q_T} -2{\bf Q}_{\xi} =0$ (mod $S$ quantum numbers).
The resolution of this ambiguity is to choose the combination of 
Higgs field VEVs giving the largest value for the product. 
There is no problem in practice due to the large powers of $T$ involved.

\section{Combinatorial corrections to mass matrices \label{fac}}

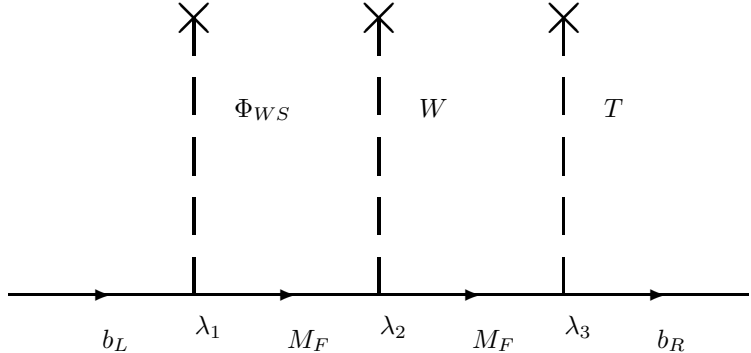
\begin{figure}
\begin{picture}(40000,13000)
\THICKLINES

\drawline\fermion[\E\REG](5000,1500)[7000]
\drawarrow[\E\ATBASE](\pmidx,\pmidy)
\global\advance \pmidy by -2000
\put(\pmidx,\pmidy){$b_L$}

\put(12000,0){$\lambda_1$}

\drawline\fermion[\E\REG](12000,1500)[7000]
\drawarrow[\E\ATBASE](\pmidx,\pmidy)
\global\advance \pmidy by -2000
\put(\pmidx,\pmidy){$M_F$}

\put(19000,0){$\lambda_2$}

\drawline\fermion[\E\REG](19000,1500)[7000]
\drawarrow[\E\ATBASE](\pmidx,\pmidy)
\global\advance \pmidy by -2000
\put(\pmidx,\pmidy){$M_F$}

\put(26000,0){$\lambda_3$}

\drawline\fermion[\E\REG](26000,1500)[7000]
\drawarrow[\E\ATBASE](\pmidx,\pmidy)
\global\advance \pmidy by -2000
\put(\pmidx,\pmidy){$b_R$}

\drawline\scalar[\N\REG](12000,1500)[5]
\global\advance \pmidx by 1500
\global\advance \pmidy by 1500
\put(\pmidx,\pmidy){$\Phi_{WS}$}
\global\advance \scalarbackx by -530
\global\advance \scalarbacky by -530
\drawline\fermion[\NE\REG](\scalarbackx,\scalarbacky)[1500]
\global\advance \scalarbacky by 1060
\drawline\fermion[\SE\REG](\scalarbackx,\scalarbacky)[1500]

\drawline\scalar[\N\REG](19000,1500)[5]
\global\advance \pmidx by 1500
\global\advance \pmidy by 1500
\put(\pmidx,\pmidy){$W$}
\global\advance \scalarbackx by -530
\global\advance \scalarbacky by -530
\drawline\fermion[\NE\REG](\scalarbackx,\scalarbacky)[1500]
\global\advance \scalarbacky by 1060
\drawline\fermion[\SE\REG](\scalarbackx,\scalarbacky)[1500]

\drawline\scalar[\N\REG](26000,1500)[5]
\global\advance \pmidx by 1500
\global\advance \pmidy by 1500
\put(\pmidx,\pmidy){$T$}
\global\advance \scalarbackx by -530
\global\advance \scalarbacky by -530
\drawline\fermion[\NE\REG](\scalarbackx,\scalarbacky)[1500]
\global\advance \scalarbacky by 1060
\drawline\fermion[\SE\REG](\scalarbackx,\scalarbacky)[1500]

\end{picture}
\caption{Possible Feynman diagram for bottom quark mass.
The crosses indicate the couplings of
the Higgs fields to the vacuum. Note that this is only 1 of 6 possible tree
level diagrams since the order of the interactions is arbitrary.}
\label{FeynDiag}
\end{figure}

We will now discuss a modification of the fitting procedure we have used
previously \cite{PreviousPapers}. The elements of the mass matrices are
determined \cite{FN} up to order of one 
factors to be a product of several Higgs vacuum 
expectation values (measured in units of 
some fundamental scale $M_F$ which we will take to be the 
Planck scale.) This is because the entries in the mass
matrices come from 
Feynman diagrams such as that shown in Fig. \ref{FeynDiag}. 
We have used such mass
matrices to fit the fermion masses and mixing angles. In fact we have managed
to produce fits with a smaller $\chi^2$ than would be expected, given the lack
of knowledge of order of 1 factors. So it is sensible to try to make a more
refined fit, which takes into account systematic factors, even though we will
still have order of 1 uncertainties which we can only average over.

One effect we have ignored up till now are the combinatorial factors due to
the internal ordering in the Feynman diagrams. That is, we have only taken
a single Feynman diagram to calculate the order of magnitude of each mass
matrix entry. What we should really do is
sum up all Feynman diagrams with given initial and final states. If we
restrict ourselves to tree level diagrams this means that we should consider
the sum over all distinct orderings of the interactions with the Higgs
VEVs. If we consider a diagram with order of magnitude
$\langle W \rangle^a \langle T \rangle^b \langle \xi \rangle^c
\langle S \rangle^d \langle \Phi_{WS} \rangle$, then the corresponding
matrix element should now be
calculated by summing over
all $\frac{(a+b+c+d+1)!}{a!b!c!d!}$ diagrams. Since in our approach only the
order of magnitude of each diagram is determined, we should add all these
diagrams together with random phases. Therefore the order of magnitude of
this matrix element is given by
$\left[ \frac{(a+b+c+d+1)!}{a!b!c!d!} \right]^{\frac{1}{2}} \langle W \rangle^a
\langle T \rangle^b \langle \xi \rangle^c \langle S \rangle^d
\langle \Phi_{WS} \rangle$.

It would seem at first sight that these combinatorial factors would greatly
modify our mass matrices and so vastly alter the fits we have already analysed
without such factors. However, this is not quite true since the factors can
to a large extent be absorbed into the Higgs VEVs. This means that we can get
similar fits, with or without such factors, but with rescaled Higgs VEVs. 
But of course there will still be some differences due to these factors, 
since they cannot be exactly absorbed by redefining the Higgs VEVs. 
One important point is that now even if we assume that the Higgs field 
$S$ has VEV $\langle S \rangle=1$, this
field still affects the fit through its contribution to the combinatorial
factors. So, whereas previously the quantum numbers 
of such a Higgs field could
be freely absorbed by other Higgs fields without changing anything, we will
now get slightly different fits depending on the precise definition of the
quantum numbers of the Higgs fields.

\section{Mass matrices and Higgs quantum numbers \label{higgs}}

The quantum numbers of the Higgs fields $S$, $W$, $T$, $\xi$ and 
$\Phi_{WS}$ were originally constructed \cite{PreviousPapers} by 
requiring them to give phenomenologically acceptable 
relations between the lepton and quark masses and their mixing angles.
As we have already emphasized, the quantum numbers of the 
$W$, $T$, $\xi$ and $\Phi_{WS}$ fields were only determined modulo 
those of the $S$ field. We here arbitrarily select a standard set 
of such Higgs field quantum numbers, with ``small'' values, and 
parameterise the equivalent sets of quantum numbers by integer parameters 
$(\alpha, \beta, \gamma, \delta)$, defining how many $S$ quantum numbers 
have been added. This leads to the  Higgs fields 
having the following charges:
\begin{eqnarray}
Q_S & = & (\frac{1}{6}, -\frac{1}{6}, 0, -1) \\
Q_W & = & (-\frac{1}{6}, -\frac{1}{3}, \frac{1}{2}, -\frac{1}{3})+\alpha Q_S \\
Q_T & = & (-\frac{1}{6}, 0, \frac{1}{6}, \frac{1}{3}) + \beta Q_S \\
Q_{\xi} & = & (\frac{1}{6}, -\frac{1}{6}, 0, 0) + \gamma Q_S \\
Q_{WS} & = & (\frac{1}{6}, \frac{1}{2}, -\frac{1}{6}, 0) + \delta Q_S
\end{eqnarray}

The orders of magnitude of the up quark, down quark and charged lepton 
mass matrices are given in terms of the Higgs fields as follows:

\begin{eqnarray}
M_U & = & H_U \Phi_{WS}^{\dagger} \\
M_D & = & H_D \Phi_{WS} \\
M_E & = & H_E \Phi_{WS}
\end{eqnarray}
where the order of magnitude of the Yukawa coupling matrix $H_U$ is 
given by:

$$
\pmatrix{
S^{1+\alpha-2\beta+2\gamma+\delta}W^{\dagger}T^2(\xi^{\dagger})^2 &
	S^{2+\alpha-2\beta-\gamma+\delta}W^{\dagger}T^2\xi &
	S^{2+\alpha-\beta-\gamma+\delta}(W^{\dagger})^2T\xi \cr
        S^{1+\alpha-2\beta+3\gamma+\delta}W^{\dagger}T^2(\xi^{\dagger})^3 &
	S^{2+\alpha-2\beta+\delta}W^{\dagger}T^2 &
	S^{2\alpha-\beta+\delta}(W^{\dagger})^2T \cr
        S^{3\gamma+\delta}(\xi^{\dagger})^3 &
	S^{1+\delta} &
	S^{-1+\alpha+\beta+\delta}W^{\dagger}T^{\dagger}\cr}
 \label{H_U}
$$
the order of magnitude of the Yukawa coupling matrix $H_D$ is given by:

$$
\pmatrix{
S^{-1-\alpha+2\beta-2\gamma-\delta}W(T^{\dagger})^2\xi^2 &
	S^{-2-\alpha+2\beta-\gamma-\delta}W(T^{\dagger})^2\xi &
	S^{2-3\beta-\gamma-\delta}T^3\xi \cr
	S^{-1-\alpha+2\beta-\gamma-\delta}W(T^{\dagger})^2\xi &
	S^{-2-\alpha+2\beta-\delta}W(T^{\dagger})^2 &
	S^{2-3\beta-\delta}T^3 \cr
	S^{-2-2\alpha+4\beta-\gamma-\delta}W^2(T^{\dagger})^4\xi &
	S^{-3-2\alpha+4\beta-\delta}W^2(T^{\dagger})^4 &
	S^{1-\alpha-\beta-\delta}WT \cr}
\label{H_D}
$$
and the order of magnitude of the Yukawa coupling matrix $H_E$ is 
given by:

$$
\pmatrix{
S^{-1-\alpha+2\beta-2\gamma-\delta}W(T^{\dagger})^2\xi^2 &
	S^{-2-\alpha+2\beta+3\gamma-\delta}W(T^{\dagger})^2(\xi^{\dagger})^3 &
        S^{2-\alpha-4\beta+\gamma-\delta}WT^4\xi^{\dagger} \cr
        S^{-1-\alpha+2\beta-5\gamma-\delta}W(T^{\dagger})^2\xi^5 &
	S^{-2-\alpha+2\beta-\delta}W(T^{\dagger})^2 &
        S^{2-\alpha-4\beta-2\gamma-\delta}WT^4\xi^2 \cr
        S^{-2-\alpha+5\beta-3\gamma-\delta}W(T^{\dagger})^5\xi^3 &
	S^{1+2\alpha-4\beta-\delta}(W^{\dagger})^2T^4 &
	S^{1-\alpha-\beta-\delta}WT \cr}
\label{H_E}
$$

The entries in the mass matrices represent the Higgs fields involved in the
Feynman diagram describing the tree-level interaction relevant for each
element. It is to be understood that the actual values of the mass matrix
elements are given by the products of the expectation values of the Higgs
fields involved. 
So, for example, $(W^{\dagger})^2$ would mean $\langle W \rangle^2$
and $S^n$ means $\langle S \rangle^{|n|}$. Also, for simplicity, we haven't
included the combinatorial factors in the mass matrices. These factors must
be included in the fit and are easily calculated from the powers of each
Higgs VEV, as in the previous section.

So the charges of the Higgs fields used in our previous paper
\cite{PreviousPapers}
were given by the choice $\alpha=1$, $\beta=1$, $\gamma=0$ and $\delta=-1$.
However, without the combinatorial factors and with $\langle S \rangle = 1$,
the choice of $\alpha$, $\beta$, $\gamma$ and $\delta$ does not affect the fit.
But now we will include these factors and
vary the integers $\alpha$, $\beta$, $\gamma$ and $\delta$ to see what
the effect on the fit is.
Clearly large values of these variables will
introduce large powers of $\langle S \rangle$ and large combinatorial factors
which will change the character of the mass matrices. Since we have derived
the model under the assumption that $\langle S \rangle \approx 1$ and that the
combinatorial factors are not very important, we should restrict ourselves to
small values of the integer parameters $\alpha$, $\beta$, $\gamma$ 
and $\delta$. Also this will satisfy our general requirement 
that the model should have ``small'' quantum numbers. We allow
$\langle S \rangle$ to vary, in order to somewhat compensate for the
combinatorial factors. We decided to allow
values -1, 0 and 1 for each parameter $\alpha$, $\beta$, $\gamma$ 
and $\delta$. This gives us a total of 81 
choices. We want to find the best fit among these 81, 
minimising a pseudo-chisquared function defined in Eq.~(\ref{e12}) below. 
To do this we first made an approximate fit for all 81 choices, giving 
an average pseudo-chisquared $\chi^2_{ave} = 2.7$, and then 
chose those fits which had $\chi^2 < 2.0$. This
left us with 14 fits, which we then analysed with higher accuracy.

\section{Results \label{results}}

As in previous papers, we fit to the experimental values given in Table
\ref{ExpMasses} and use a pseudo-chisquared function to measure how 
good a fit is. Since we are making an order of magnitude fit,  
we use the definition:
\begin{equation}
\label{e12}
\chi^2 = \sum \left[\ln \left(
\frac{m}{m_{\mbox{\small{exp}}}} \right) \right]^2
\end{equation}
where $m$ are the fitted masses and mixing angles and
$m_{\mbox{\small{exp}}}$ are the
corresponding experimental values in Table \ref{ExpMasses}. 
The $\chi^2$ was minimised by varying the Higgs VEVs, where
in this paper we also vary $\langle S \rangle$ rather than fixing
$\langle S \rangle = 1$. The Yukawa
matrices are calculated at the fundamental scale, 
which we take to be the
Planck scale. We use the first order renormalisation
group equations (RGEs) for the Standard Model 
to calculate the matrices at lower scales.
Running masses are calculated in terms of the Yukawa
couplings at 1 GeV.

\begin{table}
\caption{Experimental values of masses and mixing angles used in the fits. 
All masses are running masses at 1 GeV except the top quark mass 
which is the pole mass.}
\begin{displaymath}
\begin{array}{cc|cc}
m_u &  4 {\rm \; MeV} & m_c & 1.4 {\rm \; GeV} \\
m_d &  9 {\rm \; MeV} & m_s & 200 {\rm \; MeV} \\
m_e &  0.5 {\rm \; MeV} & m_{\mu} & 105 {\rm \; MeV} \\
M_t & 180 {\rm \; GeV} & V_{us} & 0.22 \\
m_b & 6.3 {\rm \; GeV} & V_{cb} & 0.041 \\
m_{\tau} & 1.78 {\rm \; GeV} & V_{ub} & 0.0035 \\
\end{array}
\end{displaymath}
\label{ExpMasses}
\end{table}

\begin{table}
\caption{Results for fits without any O(1) factors used.}
\begin{displaymath}
\begin{array}{rrrrlllll}

\alpha & \beta & \gamma & \delta & \langle W \rangle & \langle T \rangle &
 \langle S \rangle & \langle \xi \rangle & \chi^2 \\ \hline

-1 & -1 & -1 & 1 & 0.0672 & 0.0667 & 0.487 & 0.0331 & 1.81 \\
-1 & 0 & -1 & 1 & 0.0857 & 0.0522 & 0.33 & 0.0365 & 1.34 \\
-1 & 0 & 1 & -1 & 0.0705 & 0.0549 & 0.720 & 0.0422 & 1.88 \\
-1 & 1 & -1 & 1 & 0.0735 & 0.0525 & 0.686 & 0.0331 & 1.90 \\
-1 & 1 & 1 & 1 & 0.0857 & 0.0498 & 0.720 & 0.0402 & 1.68 \\
0 & 0 & -1 & 0 & 0.0610 & 0.0851 & 0.286 & 0.0315 & 1.47 \\
0 & 0 & -1 & 1 & 0.0671 & 0.0576 & 0.464 & 0.0402 & 1.42 \\
0 & 0 & 0 & 0 & 0.0705 & 0.0810 & 0.259 & 0.0365 & 2.04 \\
0 & 1 & 1 & 0 & 0.0945 & 0.0522 & 0.876 & 0.0365 & 1.78 \\
1 & -1 & -1 & 1 & 0.0740 & 0.0576 & 0.622 & 0.0315 & 1.91 \\
1 & -1 & 1 & 1 & 0.0777 & 0.0548 & 0.622 & 0.0402 & 2.03 \\
1 & 0 & -1 & 0 & 0.0777 & 0.0575 & 0.537 & 0.0443 & 1.87 \\
1 & 0 & -1 & 1 & 0.0740 & 0.0605 & 0.512 & 0.0331 & 1.94 \\
1 & 1 & -1 & 0 & 0.0992 & 0.0548 & 0.271 & 0.0489 & 1.68 \\

\end{array}
\end{displaymath}
\label{results1}
\end{table}

\begin{table}
\caption{Results for fits including O(1) factors.}
\begin{displaymath}
\begin{array}{rrrrlllll}

\alpha & \beta & \gamma & \delta & \langle W \rangle & \langle T \rangle &
 \langle S \rangle & \langle \xi \rangle & \chi^2 \\ \hline

-1 & -1 & -1 & 1 & 0.0741 & 0.0635 & 0.487 & 0.0331 & 1.57 \\
-1 & 0 & -1 & 1 & 0.0945 & 0.0522 & 0.347 & 0.0331 & 1.41 \\
-1 & 0 & 1 & -1 & 0.0857 & 0.0522 & 0.686 & 0.0365 & 1.59 \\
-1 & 1 & -1 & 1 & 0.0894 & 0.0525 & 0.756 & 0.0247 & 1.26 \\
-1 & 1 & 1 & 1 & 0.0945 & 0.0474 & 0.653 & 0.0365 & 1.46 \\
0 & 0 & -1 & 0 & 0.0741 & 0.0810 & 0.286 & 0.0300 & 1.27 \\
0 & 0 & -1 & 1 & 0.0857 & 0.0548 & 0.442 & 0.0347 & 1.40 \\
0 & 0 & 0 & 0 & 0.0816 & 0.0735 & 0.299 & 0.0331 & 1.37 \\
0 & 1 & 1 & 0 & 0.0945 & 0.0522 & 0.721 & 0.0331 & 1.62 \\
1 & -1 & -1 & 1 & 0.0857 & 0.0522 & 0.622 & 0.0300 & 1.44 \\
1 & -1 & 1 & 1 & 0.0900 & 0.0497 & 0.622 & 0.0383 & 1.70 \\
1 & 0 & -1 & 0 & 0.0900 & 0.0522 & 0.537 & 0.0422 & 1.79 \\
1 & 0 & -1 & 1 & 0.0816 & 0.0549 & 0.538 & 0.0331 & 1.64 \\
1 & 1 & -1 & 0 & 0.1042 & 0.0522 & 0.346 & 0.0444 & 1.85 \\

\end{array}
\end{displaymath}
\label{resultsO1}
\end{table}

\begin{table}
\caption{Typical fit without averaging over O(1) factors with $\alpha=-1$,
$\beta=1$, $\gamma=1$ and $\delta=1$. All masses are running
masses at 1 GeV except the top quark mass which is the pole mass.}
\begin{displaymath}
\begin{array}{ccc}
 & {\rm Fitted} & {\rm Experimental} \\ \hline
m_u & 4.2 {\rm \; MeV} & 4 {\rm \; MeV} \\
m_d & 4.7 {\rm \; MeV} & 9 {\rm \; MeV} \\
m_e & 0.98 {\rm \; MeV} & 0.5 {\rm \; MeV} \\
m_c & 1.22 {\rm \; GeV} & 1.4 {\rm \; GeV} \\
m_s & 340 {\rm \; MeV} & 200 {\rm \; MeV} \\
m_{\mu} & 83 {\rm \; MeV} & 105 {\rm \; MeV} \\
M_t & 220 {\rm \; GeV} & 180 {\rm \; GeV} \\
m_b & 7.2 {\rm \; GeV} & 6.3 {\rm \; GeV} \\
m_{\tau} & 1.17 {\rm \; GeV} & 1.78 {\rm \; GeV} \\
V_{us} & 0.15 & 0.22 \\
V_{cb} & 0.031 & 0.041 \\
V_{ub} & 0.0040 & 0.0035 \\
J_{CP} & 9.4 \times 10^{-6} &  2-3.5 \times 10^{-5}\\
\chi^2 & 1.68 & 
\end{array}
\end{displaymath}
\label{typical1}
\end{table}

\begin{table}
\caption{Typical fit including averaging over O(1) factors with $\alpha=-1$,
$\beta=1$, $\gamma=1$ and $\delta=1$. All masses are running
masses at 1 GeV except the top quark mass which is the pole mass.}
\begin{displaymath}
\begin{array}{ccc}
 & {\rm Fitted} & {\rm Experimental} \\ \hline
m_u & 3.1 {\rm \; MeV} & 4 {\rm \; MeV} \\
m_d & 6.6 {\rm \; MeV} & 9 {\rm \; MeV} \\
m_e & 0.76 {\rm \; MeV} & 0.5 {\rm \; MeV} \\
m_c & 1.29 {\rm \; GeV} & 1.4 {\rm \; GeV} \\
m_s & 390 {\rm \; MeV} & 200 {\rm \; MeV} \\
m_{\mu} & 85 {\rm \; MeV} & 105 {\rm \; MeV} \\
M_t & 179 {\rm \; GeV} & 180 {\rm \; GeV} \\
m_b & 7.8 {\rm \; GeV} & 6.3 {\rm \; GeV} \\
m_{\tau} & 1.29 {\rm \; GeV} & 1.78 {\rm \; GeV} \\
V_{us} & 0.21 & 0.22 \\
V_{cb} & 0.023 & 0.041 \\
V_{ub} & 0.0050 & 0.0035 \\
J_{CP} & 1.04 \times 10^{-5} & 2-3.5 \times 10^{-5}\\
\chi^2 & 1.46 & 
\end{array}
\end{displaymath}
\label{typicalO1}
\end{table}

In this section we will comment on the fits in Tables \ref{results1} and
\ref{resultsO1}, in order to highlight the 
general features of different choices for the
charges of the Higgs fields. We will start by making a comparison between
using random complex O(1) factors in the fitting procedure or 
just random phase factors.

As we have repeatedly stressed, we are only assuming some knowledge of the
order of magnitude of the mass matrix elements. This means that, when
calculating the eigenvalues of these matrices, we should at least consider
each element to have an arbitrary phase. We do this by averaging over the
calculated eigenvalues, using many random choices of phases for 
all the elements.
This avoids any accidental cancellations between quantities of the same
order of magnitude. However, we can also consider introducing random factors
which are of order 1, since this will not change the order of magnitude of
each element. Now, of course, we must decide more precisely what we mean by
an O(1) factor and how to average over the random variations. There is no
unique way to decide how large a number should be before we no longer consider
it to be of order 1. However, a reasonable choice is to say that a real
number is of order 1 if its natural logarithm lies between -1 and 1. So we
will pick random O(1) factors by taking the exponential of a number 
picked from a Gaussian distribution of width 1. 
We will now compare the results from fitting after 
averaging over these O(1) factors to the results without
using any O(1) factors (but still averaging over random phases.) Tables
\ref{results1} and \ref{resultsO1} give the $\chi^2$ values for the best 14
fits with and without O(1) factors. However, in order to illustrate in
detail the differences, we have shown the fitted masses and mixing angles for
a typical fit in Tables \ref{typical1} and \ref{typicalO1}.

The most obvious differences between using O(1) factors or not is in the 
ratio between up and down quark masses. The difference between these
masses is always increased when O(1) factors are used. The reason for this
is that the down mass is produced by two different combinations of matrix
elements. Since the down mass will typically be given by the root mean square
average of these two combinations, random factors will generally increase
this mass more than other masses which are essentially only determined by
one combination of matrix elements. In most of the fits
this helps but, in a few, the ratio is already large enough without the O(1)
factors and this makes the fit worse.
The O(1) factors also seem to
allow a better fit for the electron mass, which is too high without the
O(1) factors. This is because the down mass can still be fitted with a
lower electron mass (and so lower up mass), if the ratio of down:up masses
is increased.
It can be argued that we could adjust the spread
of O(1) factors to tune this ratio. Then this would introduce an extra
parameter in our fit, since there is no good way to fix it. However, we
would argue that we have chosen a `natural' spread of these factors and that
this spread can be considered, loosely speaking, as a definition of what we
mean by O(1).

The other effects of using O(1) factors are less important.
In the third generation the O(1) factors reduce the top mass and
increase the bottom mass, while leaving the tau mass almost unchanged.
However, these are small effects although this does usually worsen the
predictions.
In the second generation the O(1) factors increase the charm mass,
which is still too low, but also increase the strange mass, which is 
always too high in our model. The muon mass is not changed much.
These are again small effects and the improvement of the charm mass 
approximately cancels the worsening of the strange mass in the fit. 
However, this obviously increases the dominance of the contribution 
of the strange quark mass to the $\chi^2$ for the fits.
The O(1) factors increase $V_{us}$ which is still predicted too small.
They decrease $V_{cb}$ and this is usually bad but sometimes it is too high
without O(1) factors. $V_{ub}$ is increased, which sometimes helps, 
but often it is predicted too small without O(1) factors 
and too large with O(1) factors so
that, by chance, the O(1) factors make almost no difference to the $V_{ub}$
contribution to the value of $\chi^2$. Overall the O(1) factors lead to a
better prediction for the mixing angles and a small increase in CP
violation.
In general the O(1) factors improve the fit, mainly due to the better
fitting of the first generation masses and also the mixing angles
$V_{us}$ and $V_{ub}$. Of the 14 fits, only 2 are worse with the O(1) factors.
In both cases this is because the down:up mass ratio is already large enough
without O(1) factors. For $\alpha=-1$, $\beta=0$, $\gamma=-1$ and $\delta=1$
the fit
is only slightly worse with O(1) factors and is still a good fit, since the
worse up mass is compensated for by a better electron mass and $V_{us}$. For
$\alpha=1$, $\beta=1$, $\gamma=-1$ and $\delta=0$ many of the general
comments do not
apply. Here the down:up mass ratio is surprisingly smaller with O(1) factors
and the mixing angles $V_{us}$
and $V_{ub}$ are almost unchanged. This leads to a worse fit with O(1) 
factors, though we don't know why the down:up mass ratio changes in 
this way for this case.

It is much harder to give any general comparison between the
different choices of $\alpha$, $\beta$, $\gamma$ and $\delta$. The values
shown in Tables \ref{results1} and \ref{resultsO1} have been chosen since
we expected reasonably good fits from
them. They are, in fact, the best fits out of all 81 combinations of
$\alpha$, $\beta$, $\gamma$ and $\delta$ with values -1, 0 or 1. On
general grounds we would
expect worse fits from larger values of
$\alpha$, $\beta$, $\gamma$ and $\delta$. This is because large powers of
the Higgs field $\langle S \rangle$ would be required for
many matrix elements. This would then suppress the elements
unless $\langle S \rangle=1$. We would also 
expect large contributions from the factorial factors, 
but it would be unlikely that these could be
balanced by the suppression due to $\langle S \rangle$ for
all matrix elements. Therefore we
would end up with some elements larger than expected and others smaller
than expected. Since these effects have not been taken into account in our
derivation of the model, they would almost certainly spoil our assumed
relations. We will now just consider the cases with small values of
$\alpha$, $\beta$, $\gamma$ and $\delta$ that have been displayed in
the Tables \ref{results1} and \ref{resultsO1}. We have displayed the fitted
$\chi^2$ for all the fits. Tables \ref{typical1} and \ref{typicalO1} show
the fitted masses and mixing angles for typical cases, with and without
averaging over O(1) factors. Also shown is the predicted Jarlskog CP 
violation invariant $J_{CP}$ \cite{jarlskog}.

We can see that
generally the largest contributions to $\chi^2$ come from the
strange mass, up mass, electron mass, top mass and 
$V_{us}$. The strange mass is always fitted too high and is fitted the worst.
It doesn't vary much among the fits, indicating that the fit minimises the
strange mass since it is the largest single contribution to
$\chi^2$. Therefore the results clearly show that the strange mass
cannot be accurately fitted within our model, no matter what values of
$\alpha$, $\beta$, $\gamma$ and $\delta$ are chosen. So we are left with
the conclusion
that the strange mass must be accidentally small, if our model is to be
believed. This is not necessarily a problem, since we only claim to fit
order of magnitudes and a factor of 2 is not so large.

We can see some trends in the way other masses and mixing angles are
fitted with different values of $\alpha$, $\beta$, $\gamma$ and $\delta$.
There is generally a compromise between the up and electron
masses (up too low and electron too high). The down mass is then largely
determined by the ratio to the up mass, which varies between fits.
This ratio depends mainly on $\gamma$---certainly it seems that $\gamma=-1$
gives a
better ratio than $\gamma=1$. This is presumably the main reason why most of
the best fits (these 14 were the best out of 81) have $\gamma=-1$ and in fact
the two very best fits both  have $\gamma=-1$. 
Unfortunately there doesn't seem
to be any simple dependence on $\alpha$, $\beta$, $\gamma$ or $\delta$ for
the predicted
top mass or $V_{us}$. Since many of the fits are almost equally good, there is
probably some `unpredictable' dependence on the factorials which makes it
harder to determine such a dependence.
Since so
many fits are fairly good, it seems that we have little chance of really
choosing one scheme as the best. This is perhaps to be expected, since the
choices of $\alpha$, $\beta$, $\gamma$ and $\delta$ are in some sense just
a fine-tuning
of the general model. The model was `derived' to include important relations
between observed masses and mixing angles. Therefore, provided these variations
don't spoil such assumed relations, we would not expect great differences in
the overall fit. This is perhaps most obvious in that the strange mass is
fitted `equally badly' in the different variations of the model.

\section{Making order unity concept more precise \label{unity}}

We shall now take into account how to define ``order of unity'' 
more precisely, when the number of particle species, or the 
dimension of some representation to which they belong, is so
large that we do not want to consider it of ``order unity''.
We shall, however, formally proceed in the estimate below as if the
number of lines etc. in 
a Feynman diagram can be considered of 
``order unity''; in particular we shall not consider the number 
of external lines to be more than of order unity.

In the crudest approximation the following assumption should make sense: 
\newenvironment{descit}[1]{\begin{quote}
{\em #1\/}:}{\end{quote}}
\begin{descit}{Assumption A}
At the Planck scale 
all masses and couplings are of order unity and particles 
with all  required
quantum numbers---namely all quantum numbers possible---do exist.
\end{descit} 
However we should like to work to a level of 
accuracy where, for instance, the number of colours $N_c$---which 
is say 3---can no longer be counted as of order unity, 
but rather should be considered much bigger than unity.
It is then not a priori obvious whether the above crude ``order unity'' 
approximation can even be made consistent to such an accuracy 
nor, more precisely, how it should be done.

We here propose the following criterion for a successful more 
precise specification of the order unity of couplings concept:
\begin{descit}{The criterion}
The main guideline we shall use is the requirement that 
composing two order unity amplitudes, say one 
N-point function and one P-point function to one
(N+P-2)-point function by combining 
via a propagator, should result in an order unity
composite amplitude.
This propagator is here assumed to only include particles
that are not mass protected, so that they have Planck order of 
magnitude masses.
\end{descit}

The requirement that we only include propagators for 
particles which are not mass protected 
reflects the fact that, in assumption A, 
we ``assumed the order unity to be
valid for the particles at the Planck scale''. 
It means that the quarks or 
leptons themselves are not supposed to be included 
as propagator particles  
in this criterion. What we calculate are amplitudes 
due only to exchanges of 
heavy (Planck scale) particles, and all the quark or lepton propagators
must be inserted explicitly later (after the use of the 
criterion and the construction 
of the Planck scale particle exchange amplitudes). An example of such 
a later insertion can be found in subsection \ref{s6.4}.
 
Let us argue now for why we should take this criterion  
to be obeyed  as a test
for a successful implementation of assumption A:

Consider an $(N+P-2)$-point function. In tree diagram approximation we 
can ask for the contributions to this $(N+P-2)$-point function coming from 
diagrams with a propagator in a certain channel specified by, say,
$N-1$ of the external lines. For each combination of $N-1$ lines out of the 
$N+P-2$ external lines we have a contribution. The statement 
``particles with all required quantum numbers do exist'' may be
taken to suggest that none of these contributions are allowed to be much 
smaller than the other ones (i.e. we take them all to be 
of the same order of magnitude). Now such a contribution associated with a 
certain channel, in the sense of a selection of $N-1$ of the external 
lines, can be written as a $P$-point function connected by 
a propagator to an $N$-point function. So such a single contribution
is given by $N$-point and $P$-point functions with a propagator 
connecting them. 
Since we take all the contributions equal
in size, order of magnitudewise, and since there are in principle
``only a few'' (because we assumed that the $N$ etc.~of the $N$-point 
function is not large) they must all---order of 
magnitudewise---coincide with the full amplitude, 
the $(N+P-2)$-point amplitude. 
But that means that we argued for the criterion to be
fulfilled, in order that the assumption A can
be said to be true.

In the above argument we used the approximation---which is perhaps 
not so good in practice---that the
number of external particles in the considered amplitudes were of order unity 
with sufficient accuracy. For the application to the quark and lepton mass
matrices, this number is equal to the number of Higgs fields used 
plus 2 for the quarks or leptons.
However, we could say that the ``factorial corrections'' treated at length in 
section 3---the main new feature in the fits presented 
in this article---were precisely to take into account
factors arising from the number $N$ for these 
amplitudes not being of order unity 
accurately enough. Actually it turned out that the effects we studied 
were the $\sqrt{N!}$ factors, coming from 
including diagrams with the internal
channels used being changed corresponding to the 
permutation of the external lines
(or rather some of them). 

 We shall now explain how reasonable ``order unity'' $N$-point amplitudes 
are proposed, and subsequently confronted with our criterion. First
we shall, for simplicity, discuss the case of there being no conservation laws,
so that there are no forbidden couplings.

\subsection{Case of no conservation laws \label{s6.1}}
 
In order to define what the ``order unity'' should mean for the 
$N$-point amplitude, it is convenient to consider Euclidean momenta,
since then the poles in the amplitude are avoided. 
The natural choice of definition is then that the $N$-point function 
for an arbitrary set of particles, with Euclidean momenta of order 
unity in fundamental units, shall be of order unity 
in the simple mathematical sense:
\begin{equation}
A( p_1,p_2,...,p_N) \approx 1 \quad \mbox{(in Planck units)}.
\label{ais1}
\end{equation}
 
This is indeed what we shall propose, but there are a few things to have 
in mind here:

In usual notations, it is customary to use propagators for which the 
residues of ``the'' pole are normalized. But 
the propagator is the inverse---as a matrix---of the two-point
function, and if we thus use our specification, Eq.~(\ref{ais1}), to
determine even the two-point functions, we loose the freedom to
normalize the residues. 

Also we shall have in mind that there are many species of 
particles of the Planck scale mass
type (we here do not include the mass protected quarks and leptons)
and that the two-point functions make up a matrix, the inverse of which
is the propagator, which also is a matrix. 
Our postulate, Eq.~(\ref{ais1}), 
means that the individual matrix elements in the two-point function 
are just of order unity. Thus the species, in 
terms of which we express the order unity 
amplitudes are NOT mass eigenstates\footnote{If wanted, 
the mass eigenstates would have to be found by dividing 
the mass term, as a matrix, by the kinetic term, 
also a matrix, and then diagonalizing the resulting matrix.}. 
We shall make these considerations 
of the order unity calculations more precise statistically, 
by assuming that we have a statistical ensemble of 
order unity coefficients in the 
Lagrangian written in terms of fields for the various 
species, with a spread of order unity 
(in Planck units) so that
they are practically always of order unity. Since the 
free terms---the kinetic term as well as the mass term---are 
then random, these species end
up being rather random compared to the mass eigenstates of course.  

The propagator  
is then easily estimated to have matrix elements of the order 
1/(\# of species). This may be seen by arguing that the eigenvalues
of the inverse propagator are of the order $\sqrt{\# \ \mbox{of species}}$, 
while the eigenvalues of the propagator must of course then be of order 
$1/\sqrt{(\# \ \mbox{of species})}$. So the propagator is a 
factor of (\# of species)
smaller than the inverse propagator. 

Provided we use the propagator as the inverse
of the two-point function matrix, it does not really matter if we 
multiplied all the $N$ point functions by the $N$th power of any 
number (e.g.~the square root of the number of 
species $\sqrt{\# \ \mbox{of species}}$). This is 
because such a factor for each line would be cancelled by the 
propagators, or in the case of external lines by the wave function 
(re)normalization. So actually we could equivalently have postulated
the ``order unity'' rule as:
\begin{equation}
A( p_1,p_2,...,p_N) \approx 1/(\# \ \mbox{of species})^{(N/2)}
\quad \mbox{(in Planck units)}
\label{aisn1}
\end{equation}
This really may also be described as 
saying that the sum of the numerical squares 
of $A$ over all combinations of external particles should be of the order 
of unity. 

But, before we accept one of these (essentially equivalent) ans\"{a}tze, 
we should check that the criterion is fullfilled:

In order to check the criterion, we write down the double 
sum\footnote{It should be remembered here that 
our species were neither mass nor kinetic term coefficient eigenstates
so that a DOUBLE sum is needed, the propagator being a matrix.} 
over the species at the end-points of the propagator connecting the 
$N$-point and $P$-point functions to form a 
composite (N+P-2)-function. Now, since there 
are ($\# \ \mbox{of species}$) terms in
each of the two summations and we take each term to be a 
complex random number, these sums effectively each function as a factor 
of $\sqrt{\# \ \mbox{of species}}$. 
Since the matrix elements of the propagator
are of the order ($1/\# \ \mbox{of species}$) in the notation 
of Eq.~(\ref{ais1}) and unity in the notation of Eq.~(\ref{aisn1}), 
we end up in both cases 
with a composite amplitude of the correct ``order of unity''.

\subsection{ Case with e.g. $SU(N_c)$ symmetry \label{s6.2}}

In the case where there is a symmetry, like colour symmetry 
under the group $SU(N_c)$ say, we cannot simply use the rule
for the case of no conservation laws given above, because 
a random N-point function would of course almost certainly 
violate the symmetry. 
Instead let us make the ``order unity'' assumption 
for N-point functions, by defining a probability 
distribution for amplitudes. Then we can say, in a precise 
sense, that an amplitude is of ``order unity ''  
when it is of a type that can be gotten with high probablity 
from this distribution. The probability distribution is then 
specified as we describe below.

We first write down the cartesian product of
the states of the N external particles counted as incoming
(outgoing particles replaced by incoming antiparticles): 
\begin{equation}
|sp1, c1>|sp2,c2>\cdots |spN,cN> \quad \in \quad {\cal H}_N.
\label{Cartesian}
\end{equation}
Here the symbols ``$sp1$'' ... ``$spN$'' stand 
for the species (=``flavour'') of the external particles, 
while ``$c1$''...``$cN$''
denote their colour states.
Often one imagines working with (external) particle 
states $|sp1, c1>, ...|spN,cN>$ that 
have, for instance, a definite weak isospin component along
a certain quantization axis or definite values of some analogous 
colour Cartan algebra generators. However, for the
purpose of formulating 
the statistical rules defining the 
``order of unity'', we shall here rather assume 
that the external states are
\underline{random} superpositions of various, 
e.g.~colour and flavour, components.
W.r.t.  ``flavours'' (meant in the very general sense 
in which it is used here), it does 
not matter much to have this in mind, but for the 
conserved colour we risk to get many
zeroes alternating with too big numbers  for the 
amplitudes $A_N$, if we do not work
with random states. So the following formulae are 
only meant to be valid for states that are 
random superpositions! 
An N-point function 
conserving colour is given by 
specifying a bra-vector $<R_N|$, belonging to the 
colour singlet subspace of the cartesian product 
Hilbert space ${\cal H}_N$ (see Eq.~(\ref{Cartesian})), as the 
overlap
\begin{eqnarray}
A_N(p_1,sp1; p_2, sp2; ...;p_N,spN) 
\quad = \qquad & &\nonumber\\ 
<R_N(p_1,p_2,...,p_N)||sp1,c1>|sp2,c2>\cdots |spN,cN>.& &
\label{ANdef}
\end{eqnarray}
We now make a random ``order unity'' N-point function 
by choosing this bra-vector $<R_N|$ randomly. We decide to take 
a ``rotational invariant''
direction distribution for it and 
a norm of the order of the square root 
of the dimension of the space of singlets:
\begin{equation}
||<R_N|| \approx \sqrt{dim(\mbox{space of singlets})}
\label{RNdef}
\end{equation}

This latter choice may a priori look rather arbitrary, but note that 
it is just the choice that corresponds to the overlap size of $<R_N|$
with any normalized singlet state $|sing>$ in ${\cal H}_N$ becoming  
of order unity:
\begin{equation}
< |<R_N|sing>|^2 >_{R_N-averaging} \quad \approx \quad 1
\label{p}
\end{equation} 
 
Really this statistical definition of the order unity amplitudes 
is a bit superfluous and we could more simply just state that 
the amplitude $A_N$ becomes of the order of
\begin{eqnarray}
A_N(p_1,sp1; p_2, sp2; ...;p_N,spN) \quad = 
\qquad & & \nonumber\\  
\sum_{singlet_N } ``O(1)-factor \mbox{''} 
<singlet_N||sp1,c1>|sp2,c2>\cdots |spN,cN>.& &
\label{e19}
\end{eqnarray}
where we sum over a basis set of singlet states. 
We should here comment on the normalisation of the states w.r.t.~colour 
and flavour (i.e.~species). We are using the normalisation of 
Eq.~(\ref{ais1}) for the species degree of freedom. However we have 
chosen to use the normalisation of Eq.~(\ref{aisn1}) for colour in 
the sense that, if colour conservation were broken without 
changing their order of magnitude, the amplitudes would be smeared 
out so that the colours would function as species with the 
normalisation of Eq.~(\ref{aisn1}). In the case of no (colour) 
conservation laws discussed in subsection \ref{s6.1}, the propagator 
with the notation of Eq.~(\ref{aisn1}) would be just of order unity. 
So it is expected that the colour part of the propagator should 
on the average be of order unity. Indeed in the case of a colour 
conservation law, as considered in this subsection, we can easily 
estimate the propagator as the inverse of the two-point function $A_2$.
With external states belonging to a colour representation 
$\underline{a}$ of dimension $n=dim(\underline{a})$ and
having colour components $\vec{m}$ and $\vec{m'}$ respectively, 
we obtain:
\begin{eqnarray}
\lefteqn{\left(<\vec{m},spN|A_2|\vec{m'},spP'>\right)^{-1}}\\ 
& &  \approx \left(O(1)_{spN,spP'}
<\underline{1},\vec{0}|\underline{a},\vec{m};\underline{\bar{a}},-\vec{m'}>
\right)^{-1}\\
& & \approx \left(\frac{O(1)_{spN,spP'}}{\sqrt{dim (\underline{a})}}
\delta_{\vec{m},\vec{m'}} \right)^{-1} \\
& & \approx O(1)_{spN,spP'} \frac{\sqrt{n}}{\# \ \mbox{of species}}
\delta_{\vec{m},\vec{m'}} \\
& & \approx Prop_{\underline{a};sp N, sp P'}(p_N)
\delta_{\vec{m},\vec{m'}} 
\end{eqnarray}
where then\footnote{Of course the order unity numbers 
$ O(1)_{spN,spP'}$ depend on $spN$ and $spP'$, but they are just of 
\underline{order} unity. So we could logically just have left the 
indices out and written $O(1)$ for $ O(1)_{spN,spP'}$.}
\begin{equation}
 Prop_{\underline{a};sp N, sp P'}(p_N) \approx
 O(1)_{spN,spP'} \frac{\sqrt{n}}{\# \ \mbox{of species}}
\label{prop}
\end{equation} 

Now we want to show that our order unity amplitude definition, 
Eq.~(\ref{e19}), has the nice property of passing the test 
of being self-consistent under composition of an N-point 
function and a P-point function by propagator contraction.
We write down the composite (N+P-2)-point function formed from 
the propagator contracted N-point and P-point functions as follows:
\begin{eqnarray}
\sum_{spN, sp P', \underline{a} }A_N(p_1,sp1 ;p_2 , 
sp2;...;p_N,spN) & & \nonumber\\
Prop_{\underline{a};sp N, sp P'}(p_N) 
A_P(p'_1, sp 1';...;p'_P= p_N,sp P') & &
\label{propform}
\end{eqnarray}
Here $p_1, p_2, ...,p_{N-1}, p'_1,p'_2,...,p'_{P-1}$ 
are the four-momenta of
the external particles, and the first N-1 are 
attached to the N-point blob,
while the last---and primed ones---are attached 
to the P-point blob. 
Furthermore $sp$ stands for species or flavour 
with an analogous enumeration,
and $Prop_{\underline{a};sp N, sp P'}(p_N)$ 
is the propagator, which is 
a matrix, Eq.~(\ref{prop}), in flavour or species space. 
The momentum of the
propagator, determined from four-momentum conservation, 
is $p_N = p'_P$.
The triple sum runs over all pairs of 
species or flavours for each colour
representation $\underline{a}$ in the sum.

The ``self-consistency'' we want to show is that this expression,  
Eq.~(\ref{propform}),
equals $A_{N+P-2}(p_1,sp1; p_2,sp 2; ...;p_{N-1}, sp(N-1);p'_1, sp1'; ...;
p'_{P-1},sp(P-1)')$ order of magnitudewise, or rather that the statistical
distributions 
are roughly equal for the two expressions.

According to the property, Eq.~(\ref{p}), the (N+P-2)-point function 
is of the form $<R_{N+P-2}|sp1,c1>|sp2,c2>\cdots |sp(N-1),c(N-1)>|sp1',c1'>
\cdots |sp(P-1)',c(p-1)'>$,  with $<R_{P+N-2}|$ having projection 
on average unity (order of magnitudewise) on all the singlet states $|sing>$.
So we would be well on the way to proving the self-consistency 
condition, if we could show that the dimension of this singlet 
space were the same as that of the singlet space for the
composed amplitude of Eq.~(\ref{propform}). 
So as an introduction to prove the remarkable ``self-consistency'' 
\begin{eqnarray}
A_{N+P-2}(p_1,sp1; p_2,sp 2; ...;p_{N-1}, sp(N-1);p'_1, sp1'; ...;
p'_{P-1},sp(P-1)') =  & &\nonumber \\ 
\sum_{spN, sp P', \underline{a} }A_N(p_1,...;p_N,spN) 
Prop_{\underline{a};sp N, sp P'}(p_N) 
A_P(p'_1, ...;p'_P= p_N,sp P') & & 
\label{selfconsistency} 
\end{eqnarray}
we remark that the number of colour contractions---i.e.~of 
the singlets that can be used---is indeed the same for 
both sides of this equation.

This can be seen in the following way: Imagine the singlets 
``of the left hand side'' constructed by first coupling the 
colour representations of the $(N-1)$ and of the $(P-1)$ 
external lines of the $(N+P-2)$-point function to respectively
the $\underline{a}$ and $\underline{\bar{a}}$ 
representations, finally collecting the
singlets for the various $\underline{a}$-choices. Classified this way
it is easily seen that these singlets are in correspondence with
``those of the right hand side'', since the $\underline{a}$-classification
corresponds to counting according to the 
representation $\underline{a}$ of
the propagators. Thus we have the equation for the number of singlets:  
\begin{equation}
\# singlets_{N+P-2} = \sum_{\underline{a}} 
\# singlets_N( \underline{a}) * 
\# singlets_P(\underline{\overline{a}})
\label{20}
\end{equation}
where the representations $\underline{a}$ in the expressions 
$\# singlets_N( \underline{a})$ and 
$\# singlets_P(\underline{\overline{a}})$
just denote the representations to which 
the species, $sp N$ and $sp P'$ respectively, should belong. 

After having noticed the natural correspondence between the 
singlet components on the two sides of the to be proven equation,
Eq.~(\ref{selfconsistency}), we go over to considering a single
contribution---one of the $\# singlets_{N+P-2}$ in 
Eq.~(\ref{20})---to $A_{N+P-2}$. A contribution
from a single ``$|singlet>$'' out of the $\# singlets_{N+P-2}$, called say $s$,
to $A_{N+P-2}$ (the $s$-given representation $\underline{a}$
in the channel between the P-1 and the N-1 has $dim({\underline{a}}) =n$ ) is 
\begin{eqnarray}
\lefteqn{A^s_{N+P-2}=} \nonumber\\ 
& &  <singlet(s)|sp1,p_1,c1>|sp2,p_2,c2>
\cdots |sp(N-1),p_{N-1},c(N-1)>\cdot  \nonumber\\
& & \cdot|sp1',p_1',c1'>|sp2',p_2',c2'> 
\cdots |sp(P-1)',p_{P-1}',c(P-1)'> \\ 
& & = \sum_{\vec{m},spN,spP'} A^s_N|spN,p_N,\vec{m}>  
Prop_{\underline{a};sp N, sp P'}(p_N) \nonumber\\
 & & A^s_P\hbox{(with $p_P'$ outgoing)}|spP',p_P',\vec{m}> \label{woutg}\\
& & = \sum_{\vec{m},spN,spP'} A^s_N|\vec{m}> 
Prop_{\underline{a};sp N, sp P'}(p_N)  \sqrt{n} A^s_P|-\vec{m}> 
\label{wsqrtn}\\
& & \approx  \sum_{\vec{m},\vec{m'},spN, spP'} 
A^s_N|\vec{m}> Prop_{\underline{a};sp N, sp P'}(p_N) A^s_P|\vec{m'}>
\label{longp1}
\end{eqnarray}
The $|\vec{m}>$ and $|\vec{m'}>$ 
symbols denote that the $N$th, respectively $P$th, 
one of the external particles on the 
part of amplitude 
symbol $A^s_N|\vec{m}>$, respectively $A^s_P|\vec{m'}>$, 
is a particle with colour 
component $\vec{m}$,  respectively  $\vec{m'}$, 
while the irreducible representation of the colour degree of freedom 
is given as a consequence of the $s$-index. 
The first step, Eq.~(\ref{woutg}), consists in 
\underline{artificially} writing  
the  contribution $A^s_{N+P-2}$ from a given singlet $s$ as products 
of amplitudes $A^s_N$ and $A^s_P$, with a completeness sum over 
all colour states. We use an orthonormal basis $|\vec{m}>$, namely
$\sum_{\vec{m}} |\vec{m}><\vec{m}| = 1$. This is consistent 
with the order of unity flavour normalisation, Eq.~(\ref{ais1}), of the 
amplitudes $A^s_{N+P-2}$, $A^s_N$ and $A^s_P$. As a further elaboration, 
we introduce an artificial double summation for the flavour (species) 
indices $spN$ and $spP'$ over \# of species values. 
The statistical increase by a $\sqrt{\# \ \mbox{of species}}$-factor for 
each of the two summations is compensated by inserting the 
propagator, Eq.~(\ref{prop}). This is just repeating the result, 
obtained in subsection \ref{s6.1}, that the number of species is 
irrelevant for splitting an amplitude $A_{N+P-2}$ into 
amplitudes $A_N$ and $A_P$ connected by a propagator.

We remark that the amplitude denoted 
$A^s_P\hbox{(with $p_P'$ outgoing)}|spP',p_P',\vec{m}>$ 
in Eq.~(\ref{woutg}) is not normalised according to the 
same convention as the ``order unity'' amplitude obtained 
by just crossing our amplitude $A^s_P$. This is because 
we do not normalise the state with outgoing $p_P'$ 
according to our ``order unity'' rule. We rather use the relation
\begin{equation}
<\underline{1},\vec{0}|\underline{a},\vec{m};\underline{\bar{a}},-\vec{m}>
\approx \frac{O(1)}{\sqrt{dim (\underline{a})}}.
\label{26} 
\end{equation}
to obtain:
\begin{equation}
A^s_P\hbox{(with $p_P'$ outgoing)}|spP',p_P',\vec{m}>
\approx \sqrt{dim(\underline{a})} A^s_P |-\vec{m}>
\end{equation}
The reason for this normalisation is that  
$A^s_P\hbox{(with $p_P'$ outgoing)}|spP',p_P',\vec{m}>$ 
then appears as a subpart of the Clebsch-Gordan construction 
of the $A_{N+P-2}$ amplitude in the simplest way, i.e.~without any 
extra $\sqrt{dim(\underline{a})}$ factors. We are really just using 
Eq.~(\ref{26}) to convert the notation from outgoing to ingoing. 
So, in Eq.~(\ref{wsqrtn}), it is meant that both the $|\vec{m}>$ and 
$|-\vec{m}>$ states are counted as ingoing.
To get to the result Eq.~(\ref{longp1}), 
we use the following rule---which we often use in these random order one
treatments: we take it that the sum of a series of statistically roughly 
independent terms of the same size equals a typical term times the square root 
of the number of terms.

From Eq.~(\ref{longp1}) we get by summation over the different 
contributions, i.e.~by summation over $s$, that
\begin{equation}
A_{N+P-2} = \sum_{\underline{a},\vec{m},\vec{m'}, spN,spP'}
Prop_{\underline{a};sp N, sp P'}(p_N) 
 A_N|\vec{m}>A_P|\vec{m'}> .
\label{summed}
\end{equation} 
But this is just the self-consistency equation we wanted to test!
Therefore the order of 
magnitude unity, as we defined it for the (N+P-2)-point function, coincides 
with what you get by composing an N-point function with a P-point function, 
that are each of order unity in our sense, with an intermediate propagator 
(that is of course summed over) . This fact we take as very suggestive for 
our chosen definition to be a good one. It means that it passed the 
`self-consistency check', the criterion we gave at the beginning 
of section 6.

\subsection{Extraction of definite $\bf colour_i$ (i=1,2,3) states}

For the applications of the order unity amplitudes, it is not so nice to
only have them defined for ``random'' colour states. We think of states with 
respect to the three different $SU(3)$ groups in our AGUT-model, when
we here talk about colour states. However, for simplicity, in this section we
write formulae for only one of the ``generation corresponding'' colours. 
It would of course be useful to extract expressions for amplitudes of order 
one for states with definite colour indices.

As an example we shall take the cases of amplitude contributions that can
occur in the large $N_c$ limit, where we cannot use the $\epsilon$-symbol
in colours because it has (infinitely) many indices 
and only a finite number of external particles. Really let us, 
for our example here, imagine that to the singlet-contributions correspond
amplitudes that can be simply written by means of colour kronecker deltas 
in the fundamental representations (= triplets for SU(3)) 
$\delta_{\alpha\beta}$,
where then $\alpha$ and $\beta$ run over the $N_c$ colours. That is to say 
we take the contribution from one of the $<singlet|$ states 
to be proportional
to a product of such kronecker deltas $\delta_{\alpha\beta}$
\begin{eqnarray}
\lefteqn{<singlet|sp1,c1>|sp2,c2> \cdots |spN,cN> } \nonumber \\
&=& ``normalization\mbox{''} 
\left(\prod_i \delta_{\alpha_i \beta_i}\right) \cdot 
\mbox{product of the $c1^{\alpha_i} $ etc. }
\label{extraction1}
\end{eqnarray}
where there can then be various numbers of indices attached to the various 
external states depending on their 
representations. Here the $``normalization$''
factor is to be calculated so as to get the 
random states amplitude to come out right.
If we only care for getting the large $N_c$ factors right, we would not 
need to care for whether we symmetrise 
states with a few indices. Furthermore, in the cases of 
interest in our model, we have 
for one generation of colour all the time only 
triplets, singlets or antitriplets. So let us, as the example, 
consider that we only 
have contractions of triplets between the various external particles. 

With $<singlet|singlet>=1$ the statistical average 
of the square modulus of the expression in Eq.~(\ref{extraction1})
will be
\begin{equation}
\hbox{Average}(|<singlet|sp1,c1>|sp2,c2> \cdots |spN,cN>|^2) 
= \frac{1}{n_1 n_2 \cdots n_N } 
\end{equation} 
where $n_i$ is the dimension of the representation (of colour) to which 
the random state $|spi,ci>$ belongs. Or in other words
\begin{equation}
<singlet|sp1,c1>|sp2,c2> \cdots |spN,cN> 
\approx \frac{1}{\sqrt{n_1 n_2 \cdots n_N }}.
\end{equation}
Now putting the random states (in colour space) into the 
right-hand side of Eq.~(\ref{extraction1}) 
leads to the result 
\begin{eqnarray}
\lefteqn{``normalization\mbox{''} \left(\prod_i 
\delta_{\alpha_i\beta_i}\right) \cdot 
\hbox{product of the $c1^{\alpha_i}$ etc.}} \nonumber \\
&\approx &``normalization\mbox{''}\frac{1}{\sqrt[4]{n_1 n_2 \cdots n_N }}
\end{eqnarray}
in the case we considered.
We here used that two random triplets/$N_c$-plets, 
$c2^{\alpha}$ and $c3^{\beta}$ say, 
have components of the order of $ \frac{1}{\sqrt{N_c}}$ 
and the rule for summation 
of equal and independent terms giving a factor of the square root of their 
number. For each $\delta_{\alpha\beta}$ there are only $N_c$ 
non-zero terms to sum and thus only one factor $\sqrt{N_c}$. 
Hence we have, for example 
\begin{equation}
c2^{\alpha}\delta_{\alpha\beta}c3^{\beta} \approx 
\frac{1}{\sqrt{N_c}} = \frac{1}{\sqrt[4]{n_2 n_3}}.
\end{equation}
So we conclude that, for the consistency of Eq.~(\ref{extraction1}), 
we must take 
\begin{equation}
``normalization\mbox{''} \approx \frac{1}{\sqrt[4]{n_1n_2 \cdots n_N}}
\end{equation}
without any further $N_c$ factors.

Now the full ``of order unity'' amplitude was defined in 
Eq.~(\ref{e19}) as the sum over all the possible 
singlets. In cases where we only have singlets that can be written as 
simple kronecker delta contractions, as in the ``large $N_c$'' 
case we treated here, it follows that 
the full amplitude is obtained by summing 
up---one for each singlet---the expressions 
\begin{eqnarray}
\lefteqn{<singlet|sp1,c1>|sp2,c2> \cdots |spN,cN>} \nonumber \\
& =& \frac{\hbox{``O(1)-factor''}}{\sqrt[4]{n_1n_2 \cdots n_N}} 
\left(\prod_i \delta_{\alpha_i \beta_i}\right) \cdot 
\mbox{product of the $c1^{\alpha_i} $etc. }
\label{extraction2}
\end{eqnarray}
to get 
\begin{eqnarray}
\lefteqn{A_N(sp1,c1;sp2,c2;\cdots ;spN,cN)} \nonumber \\ 
&=&\sum_{\hbox{the singlets}}\frac{\hbox{``O(1)-factor''}} 
{\sqrt[4]{n_1n_2 \cdots n_N}}
\left(\prod_i \delta_{\alpha_i \beta_i}\right) \cdot
\mbox{product of the $c1^{\alpha_i} $etc. }.
\label{extraction3}
\end{eqnarray}

The fourth root normalisation factor 
could be distributed through Eq.~(\ref{extraction2}) 
or (\ref{extraction3}) so as to give 
a factor $\frac{1}{\sqrt{N_c}}$ following
each kronecker delta. In other words if, instead of the 
kronecker deltas above, we use the combination
\begin{equation}
\frac{\delta_{\alpha\beta}}{\sqrt{N_c}}
\end{equation}
where $N_c$ is the number of colours (for the $SU(3)$-group 
of the generation in question, of course $N_c=3$), we do not need 
the fourth root denominators.
That is to say we can write e.g.~Eq.~(\ref{extraction2}) as
\begin{eqnarray}
\lefteqn{<singlet|sp1,c1>|sp2,c2> \cdots |spN,cN>} 
\nonumber \\
& =& \hbox{``O(1)-factor''} 
\left(\prod_i \frac{\delta_{\alpha_i \beta_i}}{\sqrt{N_c}}\right) \cdot 
\mbox{product of the $c1^{\alpha_i} $etc. }
\label{extraction4}
\end{eqnarray}

\subsection{Implications of our order unity choice\label{s6.4}}

Now that we have chosen a prescription for making the order unity 
concept more precise, what does this choice imply for the possibility 
of finding large $N_c$ factors in our mass predictions? 

As an example we consider the characteristic prediction of our model: 
that if the diagonal elements dominate the mass matrices, then the 
quarks and charged lepton in the same generation 
have order of magnitudewise the same masses. 
Could this prediction be changed by, say, an $N_c$ factor?

In the model we use we have assumed that there are several Dirac 
fermions, as well as other particles, with masses of the order of the
Planck scale for all quantum number combinations we can think of. 
So, in particular, the quantum numbers of say a right-handed 
b-quark occur on several particles which are very heavy. 
This should be understood in the terminology of Weyl particles
as follows: There are several right-handed Weyl fields with the quantum
number combination of the right-handed b-quark and in addition several,
but one less, left-handed Weyl fields with this quantum number 
combination. So just one linear combination of these right-handed Weyl
fermions is left without a partner and is thus massless compared 
to the Planck scale. It only gets its mass, at the end, by being 
paired with a Weyl field having the quantum numbers of the left-handed 
b-quark and, only then, under the influence of the Weinberg-Salam 
Higgs field and other Higgs fields in our model. It thus gets a mass
much smaller than the Planck scale and this particle is the b-quark. 

We imagine now that we have integrated out the fields of mass of the 
order of the Planck scale and replaced their effects by effective 
N-point amplitudes for the lighter particles. For example the 
effective amplitude $A^{(mass)}_{2+h}(l,r,W,T,...)$ that describes
the scattering amplitude between the various particles described 
by the symbols $l,r,W,T,...$ which stand for their names and states: 
the $l$ and the $r$ are the left- and right-handed quarks or leptons, 
and the W, T, ... stand for our various Higgs fields. We shall be 
especially interested in the case of the states of the Higgs 
fields W, T, ...being those superpositions 
in which they occur in the vacuum. But we just wrote ``W, T, ...''
as an example and we can, of course, write a similar expression with 
any combination of our Higgs fields. Further $h$ is the number of 
the Higgs fields $W$, $T$, etc.~and the upper index ``$(mass)$''
means that we ignore the kinetic part, or equivalently that we 
take the fermion external momenta to zero.   

If the external states $W$, $T$,... are thought of 
as the vacuum condensate states---having of course zero four 
momentum---the $A^{(mass)}_{2+h}(l,r,W,T,...)$-amplitude 
becomes the two point function corresponding to the effective 
mass term in the Lagrangian for the Dirac particle formed from 
the Weyl particles $l$ and $r$.

If the kinetic energy term in the Lagrangian formulation
were simply $/\hspace*{-0.2cm}p$ times a unit matrix, we could 
see that this ``two''-point amplitude $A^{(mass)}_{2+h}$ 
is the mass-matrix (in colour 
space), but now the kinetic term coefficients 
are also random corresponding to being
of order unity in the sense which we defined. We could renormalize 
the fields so as to bring the kinetic term to the standard 
form or, equivalently, we could divide by the ratio of the 
kinetic term to $/\hspace*{-0.2cm}p$. Since the 
propagators $prop_R$ and $prop_L$ are defined as inverses of the 
kinetic terms, we really have to correct the mass-amplitude 
$A^{(mass)}_{2+h}$ by a factor $\frac{prop}{/\hspace*{-0.2 cm}p}$. 
We now consider the mass squared $m^2$
for the Dirac particle formed, which can be gotten from the relation 
\begin{eqnarray}
Tr\left[ A_{2+h}^{(mass)}(l,r,W,T,...) 
\frac{prop_R}{/\hspace*{-0.2 cm}p }
\bar{A}_{2+h}^{(mass)}(r,l,W,T,...) 
\frac{prop_L}{/\hspace*{-0.2 cm}p} \right] & & \nonumber \\
= dim(\mbox{representation}) \ m^2. & &
\label{traceeq}
\end{eqnarray}
Here these propagators are really just matrices approximately 
proportional to the identity $\frac{prop_R}{/\hspace*{-0.2 cm}p} 
\approx \sqrt{n}\delta_{\vec{m},\vec{m'}}$ in colour 
space, where $n$ is the number of colour 
states for the repesentation of $l$ or $r$
(assumed equal), $ n = dim(\mbox{representation})$. We assume 
that $l$ and $r$ belong to the same representation with respect 
to colour in order that the mass term shall not violate 
colour conservation. The trace, $Tr$, stands 
for the trace in the space of various--really colour--states 
in the $l$ and $r$ channels. If, by colour symmetry, they all have 
the same mass (squared) we just get the dimension of the representation 
of the $l$ or $r$ state spaces from this trace-summation, explaining 
the $dim(\mbox{representation})$ factor on the right-hand side of 
Eq.~(\ref{traceeq}).   
  
From our prescription Eq.~(\ref{e19}) for ``order unity'' amplitudes, 
we have
\begin{eqnarray}
A_{2+h}(r;l;W;...;T) \quad = 
\qquad & & \nonumber\\  
\sum_{singlet_N } ``O(1)-factor\mbox{''} 
<singlet_N||r>|l>|W>\cdots|T>.& &
\label{e19moved}
\end{eqnarray}
provided the external states are taken as ``random'' states---random
colour superpositions first of all.
Now, however, we want to consider definite 
colour states and we can make use of the
estimate, Eq.~(\ref{extraction4}), in subsection 6.3 to obtain:  
\begin{eqnarray}
\lefteqn{A_{2+h}(r;l;W;...;T) }\\
&=&\sum_{\hbox{the singlets}}\hbox{``O(1)-factor''} 
\left(\prod_i\frac{ \delta_{\alpha_i \beta_i}}{\sqrt{N_c}}\right) \cdot 
\mbox{product of the $c1^{\alpha_i} $etc. }
\label{from 63}
\end{eqnarray}
We have to imagine that, for the various Higgs fields $W$, $T$, etc., 
we have some colour contraction 
kronecker deltas {\em between different generation colours}. 
We note that all the Higgs fields in our model must conserve the 
diagonal colour group, which is identified with the QCD-colour group. 
So the typical Higgs field, like $T$ 
in our model, must take expectation values 
which are invariant under this diagonal $SU(3)$, 
although they break spontaneously the separate $SU(3)$'s
of the various generations. That is to say that,
for example, $T$ could be
triplet and anti-triplet under the 2nd and 3rd generation $SU(3)$'s 
respectively, and thus its vacuum expectation value would 
be proportional to an inter-generation
kronecker delta, having one index of generation 2 
and the other of generation 3.

In the simple case where we can neglect contributions containing 
$\epsilon$-symbols---the large $N_c$ approximation---we would simply 
get several kronecker deltas contracted in circular chains, because 
at the end there should be no free indices in a 
colourless expression such as Eq.~(\ref{traceeq}). 
Each ring of circularly contracted kronecker deltas is quickly 
reduced to just the trace of one kronecker delta
and thus to $N_c$, the number of colours, which is the same 
for all the generation colour groups. The dominant term will be the 
one corresponding to that $|singlet>$ which leads to the highest number
of circular chains, because that will give the highest number of 
$N_c$-factors.

For example for a diagonal mass matrix element, in which we have the same
generation of colour index on the $l$ and the $r$ states, 
it will pay best to get a kronecker delta directly contract $l$ and $r$; 
otherwise potential further contraction loops would miss the chance 
of existing, because the Higgs fields 
put between the $l$ and the $r$ in the contraction chains 
would already have been used.
We can use this observation to argue that the trace on the left-hand 
side of Eq.~(\ref{traceeq}) for a quark of a certain generation 
will be just bigger, by having one contraction-loop more, 
than the corresponding lepton in the same generation. 
That means then that this trace will be $N_c$ times bigger for a quark 
than for a lepton in the same generation---if their masses are 
given by the diagonal mass matrix elements 
(i.e.~except for $c$ and $t$ in our model). From the presence 
of the factor $dim(\mbox{representation})$ in Eq.~(\ref{traceeq}), 
it then follows that to this ``large $N_c$ approximation'', 
when diagonal matrix element dominate, 
the quark and charged lepton masses in the same generation 
are equally big (up to order 
unity factors which do not contain $N_c$ factors, as one
could perhaps have feared.). 
 
We have thus seen that one of the major predictions of our model---the 
order of magnitude degeneracy of masses in the same generation 
as long as they are dominated by diagonal matrix elements---is NOT 
modified by any big factor even if the colour number, $N_c=3$,
should be considered ``bigger than of order unity''. 
It follows that one can expect a higher accuracy 
than having to ignore a deviation by a factor $N_c=3$ in our predictions. 
Indeed our fits are accurate to better 
than to a factor 3.  We only showed this for the diagonal 
mass matrix elements 
and with ignored $\epsilon$'s, but it is not difficult to extend 
the argument to include the possibility of $\epsilon$'s for 
the intrageneration mass ratios.

It is somewhat complicated to calculate the colour-counting corrections 
for other matrix elements, but preliminary studies suggest that even 
for the off-diagonal elements the corrections to our fit 
are not very big.

\section{The expected uncertainty \label{deviation}}

According to our discussion of the many different permutations 
of the order of the Higgs field tadpoles in the Feynman diagrams 
contributing to a given mass matrix element,  
it could easily turn out that 
the mass matrix elements appear in our model as sums 
over rather many terms, all having in principle the same order of 
magnitude. There is also the possibility of having several fundamental 
scale particles with the same quantum numbers. 
If indeed there are many terms of the same approximate 
size, adding up with some random order of unity factors, 
we should be able to apply the central limit theorem to 
conclude that the sum, considered as a stochastic variable $z$, 
will have a Gaussian distribution 
in the complex plane (really a two-dimensional Gaussian distribution).

But now we mainly care for the order of magnitudes of the 
mass matrix elements and calculate the 
$\chi^2$-like quantity for the logarithm of the variable rather than 
for the variable itself. 
So we shall now consider what the spread in the logarithm is for a 
variable that is Gaussian distributed.
We shall assume, what because of their assumed random phases is true for the 
mass matrix elements in our model, that the average is zero. 
Then, for dimensional reasons, there is no way in which this 
fluctuation can depend on the width of the Gaussian distribution of 
the quantity itself. 

We now calculate the root mean square fluctuation of the logarithm 
of the stochastic variable $z$:
\begin{eqnarray}
& <(\log z)^2 - <\log z>^2>  & \\  = & \frac{\int \exp(-|z|^2) 
(\log |z|)^2 d^2z}{\int \exp(-|z|^2)} - 
\left(\frac{ \int \exp(-|z|^2) \log |z| d^2z}
{\int \exp(-|z|^2) d^2z}\right)^2 & \\
 = & \frac{\int_0^{\infty} \exp(-t)(1/2  * \log t)^2 dt}
{\int_0^{\infty}\exp(-t) dt}-
\left( \frac{\int_0^{\infty} \exp(-t) 1/2*\log t dt }
{\int_0^{\infty} \exp(-t) dt}\right)^2 & \\
 = & \frac{ \Gamma''(1)}{4\Gamma(1)}  - (\frac{\Gamma'(1)}{2\Gamma(1)})^2
 \quad = \quad \frac{\psi'(1)}{4} 
\quad  = \quad  0.4112 & 
\end{eqnarray}

If all the masses in nature really followed our model in a statistical way 
they would deviate in the logarithm from the predicted values 
statistically by a factor of $\sqrt{\psi'(1)/4}$ = 0.6412.
Now we make a fit with 12 - 4 = 8 degrees of freedom, and so  we expect the 
pseudo-chisquared of Eq.~(\ref{e12}) to be 8 x $\psi'(1)/4$
= 8 x 0.4112 = 3.29.
This to be expected pseudo-chisquared is of a very similar
order of magnitude to the pseudo-chisquareds which we find 
in our fits. So roughly 
our agreement is similar to the expected one. But actually we got
a somewhat better agreement than one should have expected for our 
presented fits!

We should bear in mind, however, that the fluctuation of the 
pseudo-chisquared itself is of the order of 35\%, 
so that the predicted pseudo-chisquared of
$3.3 \pm 1.2$ should be counted as only one standard deviation off if it
were measured to be 2.1. 
It is, of course, clear that if we fit 81 models and look at the ones 
with the smallest pseudo-chisquared, we should get some that have 
an anomalously low pseudo-chisquared. 
However, we investigated the distribution of 
our original 81 approximate fits and found a pseudo-chisquared 
fluctuating around 2.1 with a spread of 0.9. This means that the major 
part of these fits work too well! A possible explanation for this 
is that all these 81 fits are not really very different and should be 
considered roughly as just one fit, accidentally fitting too well 
by one standard deviation. An alternative explanation could be 
that some of our predictions are actually valid more accurately than 
in our model due, say, to some physics which our model has missed. 
For example we have the order of magnitude--but not exact--SU(5) 
mass predictions in our model. It could easily be that an exact SU(5) 
could lead to too good fits for our model. 

\section{Conclusion \label{conclusion}}

We have improved our previous AGUT model fits \cite{PreviousPapers} 
to the quark-lepton mass spectrum in a couple of ways:

First we have taken into account the number of different permutations 
of the Higgs field attachments to the Feynman diagram contributing 
to a given mass matrix element. This is a necessary 
correction that should be expected to be there. Secondly, 
in order to readjust the fit, it seemed necessary to let the 
VEV of the Higgs field $S$, which was 
previously set equal to unity, also be a fitted parameter. 
Because of the $S$ field now being also relevant, there are many variants 
of the model obtained by different choices of the gauge quantum numbers 
of the Higgs fields $W$, $T$, etc., which were previously only 
determined modulo those of $S$. 
This gives a lot of models, but they are in reality not very different.

We proposed a more careful definition of what an order one coupling 
assumption shall mean and investigated whether it could give 
rise to correction factors of order $N_c = 3$. Such large 
corrections do not in fact appear, at least in our prediction 
of intrageneration mass degeneracy (except for the top and charm 
quark masses, which are dominated by off-diagonal mass matrix
elements). Thus there appears to be no reason why we should 
not trust our naive calculations to an accuracy  
better than a factor of 3.    

So, as far as the charged quark and lepton mass matrices (neutrinos 
are a problem for several reasons \cite{Mark}) are concerned, we
managed to successfully fit the 9 masses,
3 mixing angles, and the CP-violation strength
with our AGUT-model, i.e.~with the gauge group $SMG^3\times U(1)_f$,
and five Higgs fields $W$, $T$, $\xi$, $S$ and $\phi_{WS}$. 
In addition to the usual Weinberg Salam Higgs vacuum expectation 
value---which in our model is replaced by $\phi_{WS}$---we only used 
the vacuum expectation values of the four other Higgs fields as free 
parameters in our fits. It should then be remarked though that:

1) With the slightly complicated gauge group representations of
our Higgs fields, some fitting of discrete quantum numbers could be said to
have taken place.

2) The expectation value of the Higgs field $S$ is so close to unity that
we only barely need it in the fits, and so we should hardly count its value
as a parameter. Its value is seldom fit to be less than $0.3$ and 
it never needs to be more than 1.

3) Really we have a significant amount of ambiguity in the discrete 
quantum numbers mentioned under 2) in the sense that many proposals,
deviating from each other by adding 
the quantum numbers of the field $S$ to one 
or several of the other Higgs fields, give very good fits.

The fits are so good that many of the fitting proposals mentioned under
point 3) even fit the data slightly better than could be expected.
We calculated an expectation for the $\chi^2$, defined in 
Eq.~(\ref{e12}) for the logarithmic comparison with data, from
the hypothesis that the mass matrix elements at first have Gaussian 
distributions. This expected value for $\chi^2$ is $3.3 \pm 1.2$, 
which roughly agrees with the $\chi^2$ values of our fits. For a lot of 
the 81 models considered, we actually obtained 
a somewhat smaller $\chi^2$ than this theoretically predicted one.
So we even have reasonably good 
agreement concerning the degree of accuracy of our fit and 
must conclude that the model fits the charged 
fermion mass matrices at least as well as one should expect. 

\section{Acknowledgements}

This work was supported by EU funds CHRX-CT-94-0621, INTAS 93-3316, 
INTAS RFBR-95-0567, SCI-0430-C (TSTS).
We should also like to thank the Slovenian Ministry of Science for 
supporting the Bled workshops, where some of this work was 
developed.

\end{document}